\documentstyle[12pt,axodraw]{article}

\begin{document}
\baselineskip=17pt  
\parskip=5pt
\begin{titlepage} \footskip=5in     

\title{    
\begin{flushright} \normalsize   
UK/TP 00-06 \vspace{0.1em} \\   
ISU-HET-00/6 \vspace{0.1em} \\    
ADP-00-52/T432 \vspace{0.5em} \\
November 2000 \vspace{3em} \\    
\end{flushright}     
\large\bf   
Can the {\boldmath$\Lambda\pi$} scattering phase shifts be large?}

\author{\normalsize\bf   
Jusak~Tandean$^a$\thanks{E-mail address: jtandean@pa.uky.edu}, \  
A.W.~Thomas$^b$\thanks{E-mail address: athomas@physics.adelaide.edu.au} 
\ and 
G.~Valencia$^c$\thanks{E-mail address: valencia@iastate.edu} \\
\normalsize\it $^a$Department of Physics and Astronomy, 
University of Kentucky, \\
\normalsize\it Lexington, Kentucky 40506-0055.\\
\normalsize\it $^b$Department of Physics and Mathematical 
Physics and \\
\normalsize\it Special Research Centre for the Subatomic 
Structure of Matter,\\ 
\normalsize\it University of Adelaide, 
Adelaide 5005, Australia. \\
\normalsize\it $^c$Department of Physics and Astronomy, 
Iowa State University, \\
\normalsize\it Ames, Iowa 50011.}
    
\date{}   
\maketitle 
  
\begin{abstract} 
  
Motivated by the presence of nearby thresholds in other 
baryon-meson channels with  $\,I=1$  and  $S=-1,\,$ 
we investigate whether the $\Lambda \pi$ scattering phase shifts 
at a center-of-mass energy equal to the $\Xi$ mass could be larger 
than suggested by lowest-order chiral perturbation theory. 
Within a coupled-channel $K$-matrix approach, we find that 
the $S$-wave phase shift could be as large as  $\,-7^\circ$.
    
\end{abstract}   
     
\end{titlepage}

\section{Introduction}
   
The $CP$-violating observable, $A$, in weak nonleptonic hyperon 
decays of the form  $\,B\rightarrow B^\prime\pi\,$  depends on 
the strong-rescattering phases of the final state~\cite{cpb}.    
At leading order, this asymmetry is given by   
\begin{eqnarray}    
A  \;=\;  
-\tan(\delta_{S}^{}-\delta_{P}^{})\, 
\sin(\phi_{S}^{}-\phi_{P}^{})   \;,  
\end{eqnarray}      
where  $\delta_{S}^{}$  and  $\delta_{P}^{}$  
($\phi_{S}^{}$  and  $\phi_{P}^{}$)  are 
the strong-rescattering (weak) phases in the $S$- and $P$-wave 
components, respectively, of the decay amplitude.   
Currently the HyperCP (E871) experiment at Fermilab is 
in the process of measuring this $CP$-violating observable  
through the asymmetry sum $\,A(\Lambda)+A(\Xi)\,$  
in the chain of decays   
$\,\Xi\rightarrow\Lambda\pi\rightarrow p\pi\pi\,$~\cite{exp}. 
The calculation of this observable, therefore, requires 
the knowledge of both the phase shifts for $N\pi$ scattering 
at the $\Lambda$ mass and those for $\Lambda\pi$ scattering 
at the $\Xi$ mass.  
The former phase shifts have been extracted from experiment 
(albeit with large errors)~\cite{roper}, but there is no 
experimental data for the latter.

An early calculation~\cite{nk} of the  $\Lambda\pi$  scattering 
phase shifts at  $m_\Xi^{}$  indicated that the  $S$-wave phase shift 
was large, the result being  $\,\delta_{S}^{}=-18.7^\circ\,$  
and  $\,\delta_{P}^{}=-2.7^\circ.\,$ 
If correct, this would suggest that $CP$ violation in both 
decays  $\,\Xi \rightarrow \Lambda \pi\,$ and 
$\Lambda \rightarrow p \pi\,$  could yield similar contributions 
to the measurement of E871, making a theoretical prediction harder. 
More recently, this calculation has been repeated in the 
context of heavy-baryon chiral perturbation theory,  
with very different results. At leading order, it was found in 
Ref.~\cite{lsw} that $\delta_{S}^{}=0$  and  
$\delta_{P}^{}= -1.7^\circ$. 
The implication of this result is that the $CP$-violating 
observable in E871 is probably dominated by $CP$ violation 
in~$\,\Lambda\rightarrow p \pi.\,$  The vanishing of  
$\delta_{S}^{}$  in this calculation results from the heavy-baryon 
limit. An estimate of relativistic corrections to 
the heavy-baryon result has been performed in Ref.~\cite{kamal}, 
where it was found (within a leading-order calculation 
in the relativistic theory) that\footnote{We have redone this 
estimate and obtained the same result, but with  $\delta_{S}^{}$  
having the opposite sign~\cite{dplp2}.}  
$\,\delta_{S}^{}=1.2^\circ\,$ and $\,\delta_{P}^{}=-1.7^\circ.\,$   
Leading-order (in heavy-baryon chiral perturbation theory) 
calculations of the $N\pi$~\cite{dplp} and $\Xi \pi$~\cite{tv} 
scattering phase shifts have also been carried out.   
They suggest that the smallness of $\delta_{S}^{}$ at lowest 
order in chiral perturbation theory~($\chi$PT) is mostly 
a kinematic effect associated with the small pion-momentum available 
in~$\,\Xi\rightarrow \Lambda \pi.\,$

Given the two very different results for $\delta_{S}^{}$ in 
$\Lambda \pi$ scattering, it is important to estimate the 
effect of physics not present in the leading-order $\chi$PT 
calculation. A first attempt to investigate 
this question was carried out in Ref.~\cite{dp}. 
Their approach was to look for possible resonant enhancements. 
To this effect, they considered the nearest resonance with the 
correct quantum numbers, the $\Sigma(1750)$ with  
$\,I=1$, $J^P=\frac{1}{2}^-.\,$    
Although the parameters of this resonance are not well known, 
the authors of Ref.~\cite{dp} allowed them to vary in a reasonable 
range to conclude that the contribution to $\delta_{S}^{}$ from 
this source was not more than about  
$\,0.5^\circ$.

In this paper we explore the possibility of an enhancement in 
$\delta_{S}^{}$ due to the presence of nearby thresholds in 
other baryon-meson channels with the same quantum numbers. 
In particular, we wish to check the role of the $S$-wave $\Sigma \pi$
channel, which has a threshold only 10 MeV above $m_\Xi^{}$.  
It is known from the Weinberg-Tomozawa theorem that the scattering 
length in this channel is very attractive~\cite{veit}.
To investigate this issue, we present two separate estimates. 
For the first one, we will take the point of view that 
any such effects can be parameterized by next-to-leading-order 
terms in chiral perturbation theory.  The authors of  
Ref.~\cite{kw} have recently studied the coupled-channel problem for  
the $\,S=-1$, $I=1\,$ baryon-meson system, within a certain model,  
and have parameterized their results in terms of specific values 
for the coupling constants in the heavy-baryon chiral Lagrangian 
up to the next-to-leading order (${\cal O}(p^2)$).   
We employ these values for the coupling constants in our first 
estimate. 
For our second estimate, we simply use leading-order $\chi$PT 
to derive all the amplitudes in the $\,S=-1$, $I=1\,$  
baryon-meson system and employ a $K$-matrix formalism to   
incorporate the effects of unitarity in the coupled-channel problem.

\section{${\cal O}(p^2)$ heavy-baryon chiral Lagrangian}

We write the chiral Lagrangian for the strong interaction 
of the lightest (octet) baryons up to order  $p^2$  in 
heavy-baryon $\chi$PT as the sum of two terms,  
\begin{eqnarray}   \label{L}  
{\cal L}  \;=\;  {\cal L}^{(1)} + {\cal L}^{(2)}   \;,  
\end{eqnarray}      
where the superscript refers to the chiral order.  
The first term is given by~\cite{JenMan} 
\begin{eqnarray}   \label{L1}   
{\cal L}^{(1)}  \;=\;  
\left\langle \bar{B}_v^{}\, {\rm i}v\cdot{\cal D} B_v^{} 
\right\rangle     
+ 2D \left\langle \bar{B}_v^{} S_v^\mu 
 \left\{ {\cal A}_\mu^{}, B_v^{} \right\} \right\rangle 
+ 2F \left\langle \bar{B}_v^{} S_v^\mu  
 \left[ {\cal A}_\mu^{}, B_v^{} \right] \right\rangle    \;,    
\end{eqnarray}      
with  $\,\langle\cdots\rangle\equiv{\rm Tr}(\cdots).\,$  
The second term can be written in the most general form 
as~\cite{kw,L2refs} 
\begin{eqnarray}   \label{L2}   
{\cal L}^{(2)}  &=&    
{\cal L}_{\rm rc}^{(2)}  \;+\;   
b_D^{} \left\langle \bar{B}_v^{}   
\left\{ \chi_+^{}, B_v^{} \right\} \right\rangle   
+ b_F^{} \left\langle \bar{B}_v^{}   
 \left[ \chi_+^{}, B_v^{} \right] \right\rangle     
+ b_0^{} \left\langle \chi_+^{} \right\rangle  
 \left\langle \bar{B}_v^{} B_v^{} \right\rangle  
\nonumber \\ && 
+\;  
2d_D^{} \left\langle \bar{B}_v^{}\,   
\bigl\{ (v\cdot{\cal A})^2, B_v^{} \bigr\} \right\rangle   
+ 2d_F^{} \left\langle \bar{B}_v^{}\,   
 \bigl[ (v\cdot{\cal A})^2, B_v^{} \bigr] \right\rangle   
\nonumber \\ && 
+\;  
2d_0^{} \left\langle \bar{B}_v^{} B_v^{} \right\rangle 
 \left\langle (v\cdot{\cal A})^2 \right\rangle   
+ 2d_1^{} \left\langle \bar{B}_v\, v\cdot{\cal A} \right\rangle 
 \left\langle v\cdot{\cal A}\, B_v \right\rangle     
\nonumber \\ && 
+\;  
2g_D^{} \left\langle \bar{B}_v^{}\,   
\bigl\{ \mbox{\boldmath$\cal A$}\cdot\mbox{\boldmath$\cal A$}, 
       B_v^{} \bigr\} \right\rangle   
+ 2g_F^{} \left\langle \bar{B}_v^{}\,   
 \bigl[ \mbox{\boldmath$\cal A$}\cdot\mbox{\boldmath$\cal A$}, 
       B_v^{} \bigr] \right\rangle   
\nonumber \\ && 
+\;  
2g_0^{} \left\langle \bar{B}_v^{} B_v^{} \right\rangle   
 \left\langle \mbox{\boldmath$\cal A$}\cdot\mbox{\boldmath$\cal A$} 
             \right\rangle   
+ 2g_1^{} \left\langle \bar{B}_v^{} \mbox{\boldmath$\cal A$} 
 \right\rangle 
 \cdot \left\langle \mbox{\boldmath$\cal A$} B_v^{} \right\rangle   
\nonumber \\ &&    
+\;  
2h_D^{} \left\langle \bar{B}_v^{} 
{\rm i}\mbox{\boldmath$\sigma$}\cdot 
\bigl\{ \mbox{\boldmath$\cal A$}\times\mbox{\boldmath$\cal A$}, 
       B_v^{} \bigr\}   
\right\rangle   
+ 2h_F^{} \left\langle \bar{B}_v^{} 
 {\rm i}\mbox{\boldmath$\sigma$}\cdot   
 \bigl[ \mbox{\boldmath$\cal A$}\times\mbox{\boldmath$\cal A$}, 
       B_v^{} \bigr]   
 \right\rangle   
\nonumber \\ &&    
+\;  
2h_1^{} 
\left\langle \bar{B}_v^{} 
{\rm i}\mbox{\boldmath$\sigma$}\times\mbox{\boldmath$\cal A$} 
\right\rangle 
\cdot \left\langle \mbox{\boldmath$\cal A$} B_v^{} \right\rangle   \;,   
\end{eqnarray}      
where  
\begin{eqnarray}   \label{L2,rc}  
{\cal L}_{\rm rc}^{(2)}  &=&  
{-1\over 2 m_{0}^{}} \left\langle   
\bar{B}_v^{}\, \bigl[ {\cal D}^2-(v\cdot{\cal D})^2 \bigr] B_v^{} 
- \bar{B}_v^{}\, \bigl[ S_v^\mu, S_v^\nu \bigr]   
 \bigl[ \bigl[ {\cal A}_\mu^{},{\cal A}_\nu^{} \bigr] ,B_v^{} \bigr]  
\right\rangle       
\nonumber \\ && 
-\;   
{{\rm i}D\over m_{0}^{}} \left( 
\left\langle \bar{B}_v^{}\, S_v^{}\cdot{\cal D} 
\bigl\{ v\cdot{\cal A}, B_v^{} \bigr\} \right\rangle       
+ \left\langle \bar{B}_v^{} S_v^\mu\, 
 \bigl\{ v\cdot{\cal A}, {\cal D}_\mu^{} B_v^{} \bigr\} 
 \right\rangle      
\right)   
\nonumber \\ && 
-\;   
{{\rm i}F\over m_{0}^{}} \left( 
\left\langle \bar{B}_v^{}\, S_v^{}\cdot{\cal D} 
\bigl[ v\cdot{\cal A}, B_v^{} \bigr] \right\rangle      
+ \left\langle \bar{B}_v^{}\, S_v^\mu\, 
 \bigl[ v\cdot{\cal A}, {\cal D}_\mu^{} B_v^{} \bigr]   
\right\rangle      
\right)  
- {D F\over m_{0}^{}} \left\langle \bar{B}_v^{}\, 
\bigl[ (v\cdot{\cal A})^2, B_v^{} \bigr]  
\right\rangle      
\nonumber \\ && 
-\;  
{D^2\over 2 m_{0}^{}} \left\langle \bar{B}_v^{}\, 
\bigl\{ v\cdot{\cal A},   
       \bigl\{ v\cdot{\cal A}, B_v^{} \bigr\} \bigr\}   
\right\rangle      
- {F^2\over 2 m_{0}^{}} \left\langle \bar{B}_v^{}\, 
 \bigl[ v\cdot{\cal A}, 
       \bigl[ v\cdot{\cal A}, B_v^{} \bigr] \bigr]  
\right\rangle         
\end{eqnarray}      
is the  $1/m_{0}^{}$  (leading relativistic) correction to 
the leading-order Lagrangian at order  $p^2$, with  
$m_{0}^{}$ being the octet-baryon mass in the chiral limit.  
The constants, $b,d,g$ and $h$,  are free parameters 
(in addition to the familiar $D$ and $F$) that occur at this 
order, and we will obtain their values from the model of 
Ref.~\cite{kw}. 
In these formulae,  $B_v^{}$  is the usual 3$\times$3 matrix 
containing the (velocity dependent) octet-baryon fields, 
$v$~the baryon velocity,  
$S_v^{}$  the spin operator, and   
\begin{eqnarray}   
\begin{array}{c}   \displaystyle  
{\cal A}_\mu^{}  \;=\;  
\mbox{${\rm i}\over2$}  
\left( \xi\, \partial_\mu^{}\xi^\dagger 
      - \xi^\dagger\, \partial_\mu^{}\xi \right) 
\;=\;  
{\partial_\mu^{}\varphi\over 2f}  \;+\;  {\cal O}(\varphi^3)  \;,    
\vspace{2ex} \\   \displaystyle 
\chi_+^{}  \;=\;  
\xi^\dagger M_\varphi^2\xi^\dagger + \xi M_\varphi^2\xi 
\;=\;  
2 M_\varphi^2 
- {1\over 4f^2}\, \bigl\{ \varphi, \{ \varphi,M_\varphi^2 \} \bigr\}   
\;+\;  {\cal O} \bigl( \varphi^4 \bigr)   \;,      
\end{array}   
\end{eqnarray}    
where  $\varphi$  is the  3$\times$3  matrix for the octet 
of pseudo-scalar bosons,  $f=f_\pi^{}=92.4\,\rm MeV$  is the 
pion-decay constant, and  
$\,M_\varphi^2=   
{\rm diag}\bigl(m_\pi^2,m_\pi^2,2m_K^2-m_\pi^2\bigr)\,$
the pseudo-scalar mass matrix in the isospin limit.

The total amplitude for  $\,\Lambda\pi\rightarrow\Lambda\pi\,$,   
up to order  $p^2$,  is derived from the diagrams in  
Fig.~\ref{diagrams}.\footnote{At this order there are no loop 
contributions. The latter begin at ${\cal O}(p^3)$.}    
In the center-of-mass~(CM) frame, it is given by     
\begin{eqnarray}   \label{M(Lp)}
{\cal M}_{\Lambda\pi}^{}  &=&   
{2m_\Lambda^{}\over f^2}\, \chi_{\rm f}^\dagger 
\begin{array}[t]{l}   \displaystyle  
\Biggl[  
{D^2\over 3 m_{0}^{}}   
\Biggl( m_\pi^2-\mbox{\boldmath$k$}^2
       + {\mbox{\boldmath$k$}^4\over 3 E_\pi^2} \Biggr)     
+ \Biggl( {4 b_D^{}\over 3}+4 b_0^{} \Biggr) m_\pi^2  
- \Biggl( {2 d_D^{}\over 3}+2 d_0^{} \Biggr) E_\pi^2   
\vspace{2ex} \\   \displaystyle 
\;+\;  
{D^2\over 3}\, 
{ 3(\mbox{\boldmath$k$}'\cdot\mbox{\boldmath$k$})^2
 - \mbox{\boldmath$k$}^4  
 \over  3 m_{0}^{} E_\pi^2 }  
\;+\;  
\mbox{\boldmath$k$}'\cdot\mbox{\boldmath$k$}\, \Biggl(  
{2 D^2\over 3}\, {m_\Sigma^{}-m_\Lambda^{}\over E_\pi^2}   
- {2 g_D^{}\over 3} - 2 g_0^{}  
\Biggr)  
\vspace{2ex} \\   \displaystyle 
\;+\;  
{2 D^2\over 3}\, 
{{\rm i}\mbox{\boldmath$\sigma$}\cdot\mbox{\boldmath$k$}' 
\times \mbox{\boldmath$k$}\over E_\pi^{}} 
\Biggl( 
1 + {E_\pi^{}\over m_{0}^{}}  
- { \mbox{\boldmath$k$}^2+\mbox{\boldmath$k$} 
   \cdot \mbox{\boldmath$k$}'  \over  2m_{0}^{} E_\pi^{} }  
\Biggr) 
\Biggr] \, \chi_{\rm i}^{}   \;,    
\end{array}
\end{eqnarray}   
where  $\chi_{\rm i}^{}$  ($\chi_{\rm f}^{}$)  is the Pauli 
spinor of the initial (final)  $\Lambda$, and  
{\boldmath$k$}  ({\boldmath$k$}$'$)  
is the three-momentum of the initial (final) pion.  
The partial-wave amplitudes are then extracted using 
standard techniques,\footnote{See, e.g., Ref.~\cite{pilkuhn}.}  
and one finds in the  $\,J=\frac{1}{2}\,$  channel 
\begin{eqnarray}   
f_{\Lambda\pi}^{(S)}  \;=\;  
{-m_\Lambda^{}\over 4\pi f^2\sqrt{s}} \Biggl[  
{D^2\over 3 m_{0}^{}}   
\Biggl( E_\pi^2-2 \mbox{\boldmath$k$}^2
       + {\mbox{\boldmath$k$}^4\over 3 E_\pi^2} \Biggr)     
+ \Biggl( {4 b_D^{}\over 3}+4 b_0^{} \Biggr) m_\pi^2  
- \Biggl( {2 d_D^{}\over 3}+2 d_0^{} \Biggr) E_\pi^2   
\Biggr]   \;,  
\end{eqnarray}   
\begin{eqnarray}   
f_{\Lambda\pi}^{(P)}  \;=\;  
{-\mbox{\boldmath$k$}^2 m_\Lambda^{}\over 4\pi f^2\sqrt{s}} \Biggl[  
{4 D^2\over 9 E_\pi^{}}  
\Biggl( 1 + {2 E_\pi^2-\mbox{\boldmath$k$}^2\over 2m_{0}^{} E_\pi^{}} 
       + {m_\Sigma^{}-m_\Lambda^{}\over 2 E_\pi^{}} \Biggr) 
- \frac{2}{9} \bigl( g_D^{}+3 g_0^{} \bigr)    
\Biggr]   \;.    
\label{cptres}
\end{eqnarray}   
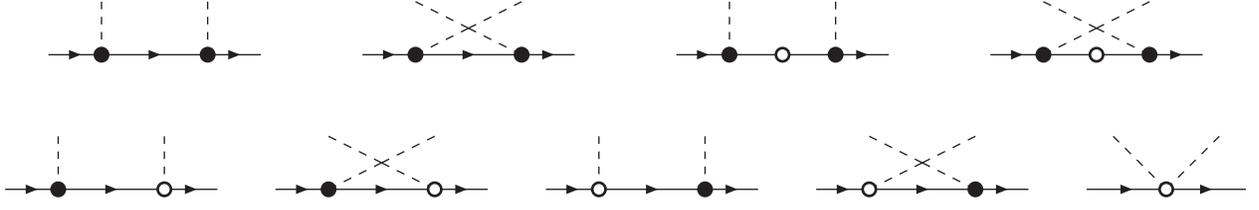
\begin{figure}[t]         
   \hspace*{\fill} 
\begin{picture}(80,50)(-40,-15)    
\ArrowLine(-40,0)(-20,0) \DashLine(-20,20)(-20,0){3} 
\ArrowLine(-20,0)(20,0) \DashLine(20,0)(20,20){3}   
\ArrowLine(20,0)(40,0) \Vertex(-20,0){3} \Vertex(20,0){3}  
\end{picture} 
   \hspace*{\fill} 
\begin{picture}(80,50)(-40,-15)    
\ArrowLine(-40,0)(-20,0) \DashLine(-20,20)(0,10){3} 
\DashLine(0,10)(20,0){3} \ArrowLine(-20,0)(20,0) 
\DashLine(-20,0)(0,10){3} \DashLine(0,10)(20,20){3}   
\ArrowLine(20,0)(40,0) \Vertex(-20,0){3} \Vertex(20,0){3}  
\end{picture}
   \hspace*{\fill} 
\begin{picture}(80,50)(-40,-15)    
\ArrowLine(-40,0)(-20,0) \DashLine(-20,20)(-20,0){3}  
\ArrowLine(-20,0)(20,0)  \DashLine(20,0)(20,20){3}   
\ArrowLine(20,0)(40,0) \Vertex(-20,0){3} \Vertex(20,0){3} 
\SetWidth{1} \BCirc(0,0){2.5}    
\end{picture}
   \hspace*{\fill} 
\begin{picture}(80,50)(-40,-15)    
\ArrowLine(-40,0)(-20,0) \DashLine(-20,20)(0,10){3} 
\DashLine(0,10)(20,0){3} \ArrowLine(-20,0)(20,0) 
\DashLine(-20,0)(0,10){3}  \DashLine(0,10)(20,20){3}   
\ArrowLine(20,0)(40,0) \Vertex(-20,0){3} \Vertex(20,0){3} 
\SetWidth{1} \BCirc(0,0){2.5}    
\end{picture}
   \hspace*{\fill} 
\\ 
   \hspace*{\fill} 
\begin{picture}(80,50)(-40,-15)    
\ArrowLine(-40,0)(-20,0) \DashLine(-20,20)(-20,0){3} 
\ArrowLine(-20,0)(20,0)  \DashLine(20,0)(20,20){3}   
\ArrowLine(20,0)(40,0) \Vertex(-20,0){3} 
\SetWidth{1} \BCirc(20,0){2.5}    
\end{picture}
   \hspace*{\fill} 
\begin{picture}(80,50)(-40,-15)    
\ArrowLine(-40,0)(-20,0) \DashLine(-20,20)(0,10){3} 
\DashLine(0,10)(20,0){3} \ArrowLine(-20,0)(20,0)   
\DashLine(-20,0)(0,10){3}  \DashLine(0,10)(20,20){3}   
\ArrowLine(20,0)(40,0) \Vertex(-20,0){3} 
\SetWidth{1} \BCirc(20,0){2.5}    
\end{picture}
   \hspace*{\fill} 
\begin{picture}(80,50)(-40,-15)    
\ArrowLine(-40,0)(-20,0) \DashLine(-20,20)(-20,0){3} 
\ArrowLine(-20,0)(20,0)  \DashLine(20,0)(20,20){3}   
\ArrowLine(20,0)(40,0) \Vertex(20,0){3} 
\SetWidth{1} \BCirc(-20,0){2.5}    
\end{picture}  
   \hspace*{\fill}   
\begin{picture}(80,50)(-40,-15)    
\ArrowLine(-40,0)(-20,0) \DashLine(-20,20)(0,10){3} 
\DashLine(0,10)(20,0){3} \ArrowLine(-20,0)(20,0)   
\DashLine(-20,0)(0,10){3}  \DashLine(0,10)(20,20){3}   
\ArrowLine(20,0)(40,0) \Vertex(20,0){3} 
\SetWidth{1} \BCirc(-20,0){2.5}    
\end{picture}  
   \hspace*{\fill} 
\begin{picture}(60,50)(-30,-15)    
\ArrowLine(-30,0)(0,0) \DashLine(-20,20)(0,0){3} 
\DashLine(0,0)(20,20){3} \ArrowLine(0,0)(30,0) 
\SetWidth{1} \BCirc(0,0){2.5}    
\end{picture}
   \hspace*{\fill} 
\caption{\label{diagrams}%
Diagrams for  $\,\Lambda\pi\rightarrow\Lambda\pi\,$   
up to order  $p^2$.  
A dashed (solid) line denotes a pion (octet baryon) field.  
The baryon in the intermediate states is~$\Sigma$.  
Solid and hollow vertices are generated by  
${\cal L}^{(1)}$  in Eq.~(\ref{L1}) and  
${\cal L}^{(2)}$  in Eq.~(\ref{L2}),  respectively.}  
\end{figure}             

For our numerical calculation, we will adopt the parameter 
values provided by Ref.~\cite{kw}.  
In that work, the chiral Lagrangian  ${\cal L}$  in Eq.~(\ref{L})  
was used as a starting point for constructing a coupled-channel 
potential model to study  
$\,N\bar{K}\rightarrow N\bar{K},\Lambda\pi,\Sigma\pi\;$
and other measured processes.  
We employ in particular the parameter values extracted in  
Ref.~\cite{RamKWW},  
\begin{eqnarray}   \label{parameters}
\begin{array}{c}   \displaystyle  
D  \;=\;  0.782   \;,   
\hspace{2em}  
m_{0}^{}=0.869\, {\rm GeV}   \;,   
\vspace{2ex} \\   \displaystyle    
b_0^{}  \;=\;  -0.320\, {\rm GeV}^{-1}   \;, 
\hspace{2em}  
b_D^{}  \;=\;  0.066\, {\rm GeV}^{-1}   \;,  
\vspace{2ex} \\   \displaystyle    
d_0^{}  \;=\;  -0.996\, {\rm GeV}^{-1}   \;,  
\hspace{2em}  
d_D^{}  \;=\;  0.512\, {\rm GeV}^{-1}   \;,  
\vspace{2ex} \\   \displaystyle    
g_0^{}  \;=\;  -1.492\, {\rm GeV}^{-1}   \;,  
\hspace{2em}  
g_D^{}  \;=\;  0.320\, {\rm GeV}^{-1}   \;.  
\end{array}   
\end{eqnarray}   
Thus, with  $\,\sqrt{s}=m_\Xi^{}\,$  and  
$\,|\mbox{\boldmath$k$}|\simeq 0.137\, {\rm GeV},\,$  
we obtain for the  $\,J=\frac{1}{2}\,$  channel the phase 
shifts 
\begin{eqnarray}   \label{deltaSP}
\delta_{S}^{}  \;\simeq\;  -2.5^\circ   \;,    
\hspace{2em}  
\delta_{P}^{}  \;\simeq\;  -3.3^\circ   \;.  
\end{eqnarray}   
Of course, the parameters in Eq.~(\ref{parameters}) are not 
known precisely. 
If we allow them to take the following ranges of values (in the 
same units as before):   
\begin{eqnarray}   
\begin{array}{c}   \displaystyle  
0.4  \;<\; D  \;<\;  0.8   \;,   
\hspace{2em}  
0.7  \;<\;  m_{0}^{}  \;<\; 1.2   \;,   
\vspace{2ex} \\   \displaystyle    
-0.6  \;<\;  b_0^{}  \;<\;  -0.3   \;, 
\hspace{2em}  
0.02  \;<\;  b_D^{}  \;<\;  0.08   \;, 
\vspace{2ex} \\   \displaystyle    
-1.0  \;<\;  d_0^{}  \;<\;  -0.7   \;,  
\hspace{2em}  
0.3  \;<\;  d_D^{}  \;<\;  0.6   \;,  
\vspace{2ex} \\   \displaystyle    
-1.5   \;<\;  g_0^{}  \;<\;  -1.0   \;,  
\hspace{2em}  
0.3  \;<\;  g_D^{}  \;<\;  0.5   \;,  
\end{array}   
\end{eqnarray}   
which are suggested by various tree- and loop-level 
$\chi$PT calculations~\cite{JenMan,various}, as well as 
the results of Ref.~\cite{kw},  we find the following ranges 
for the phase shifts:
\begin{equation}   \label{range}  
-3.0^\circ  \;<\; \delta_{S}^{}  \;<\;  +0.4^\circ   \;,  
\hspace{2em}   
-3.5^\circ  \;<\; \delta_{P}^{}  \;<\;  -1.2^\circ   \;.  
\end{equation}     
The contributions of the lowest-order terms to these numbers 
are  $\,\delta_{S}^{(1)}=0\,$  and  
$\,-1.7^\circ<\delta_{P}^{(1)}<-0.4^\circ.\,$          
The  $1/m_{0}^{}$  terms, especially in the $S$-wave, give small 
corrections,  $\,-0.06^\circ<\delta_{S}^{\rm rc}<-0.01^\circ\,$  
and  $\,-0.4^\circ<\delta_{P}^{\rm rc}<-0.1^\circ.\,$      
Therefore, the rest of the $p^2$ terms in the $S$-wave generate 
the bulk of~$\delta_{S}^{}$,  and  those in the $P$-wave are 
comparable to the lowest-order term in their contribution 
to~$\delta_{P}^{}$.

It is useful to compare the result above with the leading-order 
result of Ref.~\cite{lsw}. In that calculation, 
the spin-$\frac{3}{2}$  decuplet-baryon degrees of freedom are 
also included in the chiral Lagrangian, so that at leading 
order~\cite{JenMan}  
\begin{eqnarray}   \label{L1'}   
{\cal L}^{(1)}  &=&  
\left\langle \bar{B}_v^{}\, 
{\rm i}v\cdot{\cal D} B_v^{} \right\rangle   
+ 2 D \left\langle \bar{B}_v^{} S_v^\mu 
 \left\{ {\cal A}_\mu^{}, B_v^{} \right\} \right\rangle  
+ 2 F \left\langle \bar{B}_v^{} S_v^\mu 
 \left[ {\cal A}_\mu^{}, B_v^{} \right] \right\rangle  
\nonumber \\ && 
-\;    
\bar{T}_v^\mu\, {\rm i}v\cdot{\cal D} T_{v\mu}^{}  
+ \Delta m\, \bar{T}_v^\mu T_{v\mu}^{}  
+ {\cal C} \left( \bar{T}_v^\mu {\cal A}_\mu^{} B_v^{} 
                 + \bar{B}_v^{} {\cal A}_\mu^{} T_v^\mu \right)    
+ 2{\cal H}\; 
 \bar{T}_v^\mu\, S_v^{}\cdot{\cal A}\, T_{v\mu}^{}   \;,         
\end{eqnarray}      
where  $T_v^\mu$  represents the baryon-decuplet 
fields,\footnote{For hadronic fields, 
we follow the notation of Ref.~\cite{atv}.} 
$m_T^{}$  is the decuplet-baryon mass in the chiral limit, 
and  $\,\Delta m=m_T^{}-m_{0}^{}.\,$    
The resulting amplitude for  
$\,\Lambda\pi\rightarrow\Lambda\pi\,$  in the  $\,J=\frac{1}{2}\,$  
channel receives nonzero contributions from the last three 
diagrams in  Fig.~\ref{diagrams,p1} and is given by   
\begin{eqnarray}   
{\cal M}_{\Lambda\pi}^{}  &=&  
{2m_\Lambda^{}\over f^2}\, \chi_{\rm f}^\dagger\, 
\mbox{\boldmath$k$}\cdot \mbox{\boldmath$k$}' \left( 
{\frac{1}{3} D^2\over \sqrt{s}-m_{\Sigma}^{}}  
\,-\,  
{\frac{1}{3} D^2\over E_\pi^{}-E_\Lambda^{}+m_{\Sigma}^{}}  
\,-\,  
{ \frac{1}{6}_{}^{} {\cal C}^2  \over  
 E_\pi^{}-E_\Lambda^{}+m_{\Sigma^*}^{} }  
\right) \chi_{\rm i}^{}    
\nonumber \\ && \!\!\! 
+\;    
{2m_\Lambda^{}\over f^2}\, \chi_{\rm f}^\dagger\,   
{\rm i}\mbox{\boldmath$\sigma$} 
\cdot\mbox{\boldmath$k$}'\times\mbox{\boldmath$k$} \left(  
{\frac{1}{3} D^2\over \sqrt{s}-m_{\Sigma}^{}}  
\,+\,  
{\frac{1}{3} D^2\over E_\pi^{}-E_\Lambda^{}+m_{\Sigma}^{}}  
\,-\,  
{ \frac{1}{12}_{}^{} {\cal C}^2  \over  
 E_\pi^{}-E_\Lambda^{}+m_{\Sigma^*}^{} }  
\right) \chi_{\rm i}^{}   \;,   
\end{eqnarray}   
the  $\Sigma$  and  $\Sigma^*$  being the intermediate baryons.
This result, at order  $p^1$,  actually contains some contributions 
from  the chiral Lagrangian of order $p^2$ which are implicit in 
the denominators.  
The partial-wave amplitudes in the  $\,J=\frac{1}{2}\,$  channel 
are then   
\begin{eqnarray}   \label{lp,p1}
f_{\Lambda\pi}^{(S)}  \;=\;  0   \;,  
\hspace{2em}   
f_{\Lambda\pi}^{(P)}  \;=\;  
{-\mbox{\boldmath$k$}^2 m_\Lambda^{}\over 4\pi f^2\sqrt{s}} 
\left( 
{\frac{1}{3} D^2\over \sqrt{s}-m_{\Sigma}^{}}  
\,+\,  
{\frac{1}{9}_{}^{} D^2\over E_\pi^{}-E_\Lambda^{}+m_{\Sigma}^{}}  
\,-\,  
{\frac{1}{9} {\cal C}^2\over E_\pi^{}-E_\Lambda^{}+m_{\Sigma^*}^{}}  
\right)   \;.     
\end{eqnarray}
Consequently, using  $\,\sqrt{s}=m_\Xi^{}\,$  
and the tree-level values\footnote{These are extracted from 
hyperon semileptonic decays (which also give  $\,F=0.5$) 
and the strong decays  $\,T\rightarrow B\phi,\,$  
respectively.}  
$\,D=0.8\,$  and  $\,{\cal C}=1.7,\,$  we find  
for the  $\,J=\frac{1}{2}\,$  channel the phase shifts 
\begin{eqnarray}   
\delta_{S}^{}  \;=\;  0   \;,   
\hspace{3em}   
\delta_{P}^{}  \;\simeq\;  -1.5^\circ   \;.    
\end{eqnarray}    
If chiral-symmetric masses  
$\,m_\Sigma^{}=m_{0}^{}\simeq 1.15\,\rm GeV\,$  and  
$\,m_{\Sigma^*}^{}=m_T^{}\simeq 1.38\,\rm GeV\,$ are used instead,   
we obtain  $\,\delta_{P}^{}\simeq -1.0^\circ.\,$    
In each of these  $\delta_{P}^{}$  values,  
roughly~$\,-2^\circ\,$  arises from the two diagrams involving 
the~$\Sigma$, and about~$+0.8^\circ$ comes from the diagram 
containing the~$\Sigma^*$.  
One can see that the value of the  $\Sigma$  contribution is 
compatible with the  $\delta_{P}^{(1)}$  range quoted above.  
However, the  $\Sigma^*$ contribution is opposite in sign  
to all the ${\cal O}(p^2)$  contributions to  $\delta_{P}^{}$  
that we have estimated using the Lagrangian in  Eq.~(\ref{L2}).  
This suggests that there are additional contributions beyond 
that of the  $\Sigma^*$  which can be expected to be 
significant.

\begin{figure}[t]         
   \hspace*{\fill} 
\begin{picture}(60,40)(-30,-15)    
\ArrowLine(-30,0)(0,0) \DashLine(-20,20)(0,0){3} 
\DashLine(0,0)(20,20){3} \ArrowLine(0,0)(30,0) \Vertex(0,0){3}    
\end{picture}
   \hspace*{\fill} 
\begin{picture}(80,40)(-40,-15)    
\ArrowLine(-40,0)(-20,0) \DashLine(-20,20)(-20,0){3} 
\ArrowLine(-20,0)(20,0) \DashLine(20,0)(20,20){3}   
\ArrowLine(20,0)(40,0) \Vertex(-20,0){3} \Vertex(20,0){3}  
\end{picture} 
   \hspace*{\fill} 
\begin{picture}(80,40)(-40,-15)    
\ArrowLine(-40,0)(-20,0) \DashLine(-20,20)(0,10){3} 
\DashLine(0,10)(20,0){3} \ArrowLine(-20,0)(20,0) 
\DashLine(-20,0)(0,10){3} \DashLine(0,10)(20,20){3}   
\ArrowLine(20,0)(40,0) \Vertex(-20,0){3} \Vertex(20,0){3}  
\end{picture}
   \hspace*{\fill} 
\begin{picture}(80,40)(-40,-15)    
\ArrowLine(-40,0)(-20,0) \DashLine(-20,20)(0,10){3} 
\DashLine(0,10)(20,0){3} 
\Line(-20,1)(20,1) \Line(-20,-1)(20,-1) \ArrowLine(-1,0)(1,0) 
\DashLine(-20,0)(0,10){3} \DashLine(0,10)(20,20){3}   
\ArrowLine(20,0)(40,0) \Vertex(-20,0){3} \Vertex(20,0){3}  
\end{picture}
   \hspace*{\fill} 
\caption{\label{diagrams,p1}%
Diagrams for  $\,B\phi\rightarrow B'\phi'\,$  in 
the  $\,J=\frac{1}{2}\,$  channel at leading order in  $\chi$PT, 
including decuplet-baryon contributions.  
A dashed line denotes a meson field, and a single (double) 
solid-line denotes an octet-baryon (decuplet-baryon) field.  
Vertices are generated by  ${\cal L}^{(1)}$  in  Eq.~(\ref{L1'}).}
\end{figure}
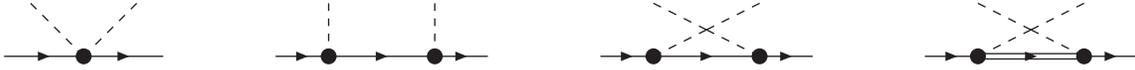             

In Ref.~\cite{kamal}, the baryons are not treated as heavy 
and the phase shifts are computed using the relativistic version 
of the lowest-order Lagrangian in Eq.~(\ref{L1'}).  
We have repeated the calculation (with  $\,D=0.8$  and    
${\cal C}=1.7$)  and found  $\,\delta_{S}^{}=-1.2^\circ\,$   
and $\,\delta_{P}^{}=-1.7^\circ,\,$   
which agrees with the result of Ref.~\cite{kamal}, 
except that  $\delta_{S}^{}$  has the opposite sign.\footnote{      
We have checked that expanding the part of the amplitude 
arising from the~$\Sigma$ diagrams up to order~$p^2$  does lead 
to the~$D^2$  terms in the heavy-baryon  ${\cal O}(p^2)$   
amplitude in  Eq.~(\ref{M(Lp)}).}       
In $\delta_{S}^{}$  here,  
only  $\,-0.1^\circ$  is generated by the  $\Sigma$-mediated 
diagrams, with the rest,  $\,-1.1^\circ,\,$  coming from 
the  $\Sigma^*$-mediated diagram.    
We can see that the~$\Sigma$ contribution is comparable to 
the~$\delta_{S}^{\rm rc}$  values in the  ${\cal O}(p^2)$  
heavy-baryon estimate.   
In contrast, the dominant $\Sigma^*$ contribution   
is roughly only half of the  $\delta_{S}^{}$  value in 
the~${\cal O}(p^2)$  heavy-baryon result, although the two 
have the same sign.       
This again suggests that other contributions in 
addition to that of  the~$\Sigma^*$  may be important.   
In the $P$-wave, the $\Sigma$ and $\Sigma^*$  
contributions to  $\delta_{P}^{}$  are  $\,-2.7^\circ$  
and  $\,+1.0^\circ,\,$  respectively,  
and so these are similar to their~${\cal O}(p^1)$  
heavy-baryon  counterparts.

\section{\boldmath$K$-matrix approach}

The SU(3) picture that we have implies that 
the  $\Lambda\pi$  state is coupled to the states  
$\,\Sigma\pi, N\bar{K}, \Sigma\eta,$ and $\Xi K\,$   
with  $\,S=-1$  and  $I=1.\,$  
Thus the $\Lambda\pi$  scattering can be treated as a problem 
with five coupled channels.   
Although at  $\sqrt{s}=m_\Xi^{}$  all the inelastic channels 
are below threshold, they may significantly affect 
the elastic one through unitarity constraints. 
The inclusion of such kinematically closed channels has been  
recently shown to be important in the case of  $N\bar{K}$ 
interactions~\cite{or}.

In order to estimate the impact on the $\Lambda\pi$ channel 
of the others coupled to it, we employ a  $K$-matrix approach.  
This method guarantees that the resulting partial-wave 
amplitudes satisfy unitarity exactly.  
We follow the formalism described in Ref.~\cite{pilkuhn}.  
For the  $K$-matrix elements, we will make the simplest 
approximation and use only the partial-wave amplitudes 
at leading order in~$\chi$PT, obtained from  
${\cal L}^{(1)}$  in Eq.~(\ref{L1'}).

The relevant partial-waves can be extracted by choosing 
the five isospin states   
\begin{eqnarray}   
\begin{array}{c}   \displaystyle      
|\Lambda\pi, I=1\rangle  \;=\;  \bigl|\Lambda\pi^-\bigr\rangle   \;,  
\hspace{3em}   
|\Sigma\pi, I=1\rangle  \;=\;  
\mbox{$1\over\sqrt{2}$}   
\bigl( \bigl| \Sigma^0\pi^- \bigr\rangle  
      - \bigl| \Sigma^-\pi^0 \bigr\rangle \bigr)   \;,  
\vspace{1ex} \\   \displaystyle 
\bigl| N\bar{K}, I=1 \bigr\rangle  \;=\;  
- \bigl| n K^- \bigr\rangle    \;,  
\hspace{2em}  
|\Sigma\eta, I=1\rangle  \;=\;  
\bigl| \Sigma^-\eta \bigr\rangle   \;,  
\hspace{2em}   
|\Xi K, I=1\rangle  \;=\;  
- \bigl| \Xi^- K^0 \bigr\rangle   \;.     
\end{array}      
\end{eqnarray}      
The phase convention here is consistent with the structure 
of the  $\varphi$  and  $B_v^{}$  matrices.

\begin{figure}[ht]         
   \hspace*{\fill}   
\begin{picture}(300,240)(105,-220)   
\LinAxis(125,0)(125,-200)(5,5,3,0,1) 
\LinAxis(300,0)(300,-200)(5,5,-3,0,1) 
\LinAxis(125,0)(300,0)(6,2,-2,0,1) 
\LinAxis(125,-200)(300,-200)(6,2,2,0,1)  
\LinAxis(125,0)(300,0)(6,2,-3,1,1) 
\LinAxis(125,-200)(300,-200)(6,2,3,1,1)  
\DashLine(148.4,0)(148.4,-200){1} 
\DashLine(178.8,0)(178.8,-200){1} 
\DashLine(268.0,0)(268.0,-200){1} 
\DashLine(289.4,0)(289.4,-200){1} 
%
%
\SetWidth{1}
\Curve{(125.863, 0)(126.733, -3.86278)(127.604, -5.61673) 
    (128.475, -7.06962)(129.345, -8.3855)(130.216, -9.62593) 
    (131.087, -10.8215)(131.957, -11.9896)(132.828, -13.1412) 
    (133.699, -14.2836)(134.569, -15.4216)(135.44, -16.5587) 
    (136.311, -17.6972)(137.181, -18.8388)(138.052, -19.9846) 
    (138.923, -21.1353)(139.794, -22.2911)(140.664, -23.4521) 
    (141.535, -24.618)(142.406, -25.7884)(143.276, -26.9627) 
    (144.147, -28.14)(145.018, -29.3194)(145.888, -30.4997) 
    (146.759, -31.6793)(147.63, -32.8559)(148.5, -34.0184) 
    (149.371, -35.2)(150.242, -36.3755)(151.112, -37.5429) 
    (151.983, -38.7002)(152.854, -39.8451)(153.725, -40.9756) 
    (154.595, -42.089)(155.466, -43.1828)(156.337, -44.2544) 
    (157.207, -45.3008)(158.078, -46.319)(158.949, -47.3057) 
    (159.819, -48.2575)(160.69, -49.1707)(161.561, -50.0412) 
    (162.431, -50.8649)(163.302, -51.6371)(164.173, -52.3527) 
    (165.043, -53.0062)(165.914, -53.5913)(166.785, -54.101) 
    (167.656, -54.5275)(168.526, -54.8617)(169.397, -55.0928) 
    (170.268, -55.2081)(171.138, -55.192)(172.009, -55.0252) 
    (172.88, -54.6823)(173.75, -54.1297)(174.621, -53.3194) 
    (175.492, -52.1788)(176.362, -50.5866)(177.233, -48.3041) 
    (178.104, -44.6853)(178.974, -35.8674)(179.845, -37.2118) 
    (180.716, -38.3835)(181.587, -39.5213)(182.457, -40.6511) 
    (183.328, -41.7837)(184.199, -42.9249)(185.069, -44.0787) 
   (185.94, -45.2479)(186.811, -46.4349)(187.681, -47.6417) 
    (188.552, -48.8703)(189.423, -50.1224)(190.293, -51.3997) 
    (191.164, -52.704)(192.035, -54.0368)(192.905, -55.4) 
    (193.776, -56.7952)(194.647, -58.2242)(195.518, -59.6885) 
    (196.388, -61.1899)(197.259, -62.7301)(198.13, -64.3109) 
    (199., -65.9337)(199.871, -67.6002)(200.742, -69.312) 
    (201.612, -71.0706)(202.483, -72.8772)(203.354, -74.7331) 
    (204.224, -76.6395)(205.095, -78.5972)(205.966, -80.6069) 
    (206.836, -82.6692)(207.707, -84.7842)(208.578, -86.9519) 
    (209.449, -89.1718)(210.319, -91.4433)(211.19, -93.7651) 
    (212.061, -96.1357)(212.931, -98.553)(213.802, -101.015) 
    (214.673, -103.517)(215.543, -106.058)(216.414, -108.633) 
    (217.285, -111.237)(218.155, -113.866)(219.026, -116.514) 
    (219.897, -119.176)(220.767, -121.846)(221.638, -124.517) 
    (222.509, -127.184)(223.38, -129.838)(224.25, -132.475) 
    (225.121, -135.086)(225.992, -137.666)(226.862, -140.207) 
    (227.733, -142.704)(228.604, -145.15)(229.474, -147.541) 
    (230.345, -149.869)(231.216, -152.132)(232.086, -154.324) 
    (232.957, -156.441)(233.828, -158.479)(234.698, -160.436) 
    (235.569, -162.309)(236.44, -164.096)(237.311, -165.795) 
    (238.181, -167.404)(239.052, -168.923)(239.923, -170.35) 
    (240.793, -171.685)(241.664, -172.928)(242.535, -174.079) 
    (243.405, -175.138)(244.276, -176.105)(245.147, -176.981) 
    (246.017, -177.765)(246.888, -178.459)(247.759, -179.063) 
    (248.629, -179.577)(249.5, -180.002)(250.371, -180.338) 
    (251.242, -180.584)(252.112, -180.741)(252.983, -180.807) 
    (253.854, -180.784)(254.724, -180.668)(255.595, -180.459) 
    (256.466, -180.154)(257.336, -179.75)(258.207, -179.243) 
    (259.078, -178.627)(259.948, -177.896)(260.819, -177.039) 
    (261.69, -176.043)(262.56, -174.89)(263.431, -173.552) 
    (264.302, -171.987)(265.173, -170.129)(266.043, -167.847) 
    (266.914, -164.843)(267.785, -159.85)(268.655, -147.415) 
    (269.526, -143.237)(270.397, -140.189)(271.267, -137.636) 
    (272.138, -135.366)(273.009, -133.272)(273.879, -131.295) 
    (274.75, -129.394)(275.621, -127.539)(276.491, -125.706) 
    (277.362, -123.878)(278.233, -122.037)(279.104, -120.165) 
    (279.974, -118.246)(280.845, -116.263)(281.716, -114.194) 
    (282.586, -112.014)(283.457, -109.693)(284.328, -107.188) 
    (285.198, -104.441)(286.069, -101.362)(286.94, -97.7975) 
    (287.81, -93.4424)(288.681, -87.4594)(289.552, -78.3439) 
    (290.422, -87.6367)(291.293, -92.9983)(292.164, -97.1721) 
    (293.035, -100.675)(293.905, -103.724)(294.776, -106.437) 
    (295.647, -108.885)(296.517, -111.117)(297.388, -113.169) 
    (298.259, -115.067)(299.129, -116.832)(300., -118.479)}
\footnotesize  
\rText(90,-100)[][l]{${\rm Re}\,\delta_{S}^{}\;(\rm degrees)$}   
\Text(212.5,-215)[t]{$\sqrt{s}\;(\rm GeV)$}   
\Text(120,0)[r]{$0$} \Text(120,-40)[r]{$-10$} 
\Text(120,-80)[r]{$-20$} \Text(120,-120)[r]{$-30$} 
\Text(120,-160)[r]{$-40$} \Text(120,-200)[r]{$-50$} 
\Text(168.8,-205)[t]{$1.4$} \Text(227.1,-205)[t]{$1.6$} 
\Text(285.4,-205)[t]{$1.8$} 
\end{picture} 
   \hspace*{\fill} 
\caption{\label{delta_S}%
$S$-wave phase shift in $\Lambda\pi$ scattering
as a function of the center-of-mass energy.  
The dotted lines mark the thresholds of the 
$\,\Sigma\pi, N\bar{K}, \Sigma\eta,$ and $\Xi K\,$   
channels, respectively. (Note that $\,m_\Xi^{}\simeq1.32\,\rm GeV\,$ 
is just 10 MeV  below the~$\Sigma\pi$ threshold.)}    
\end{figure}
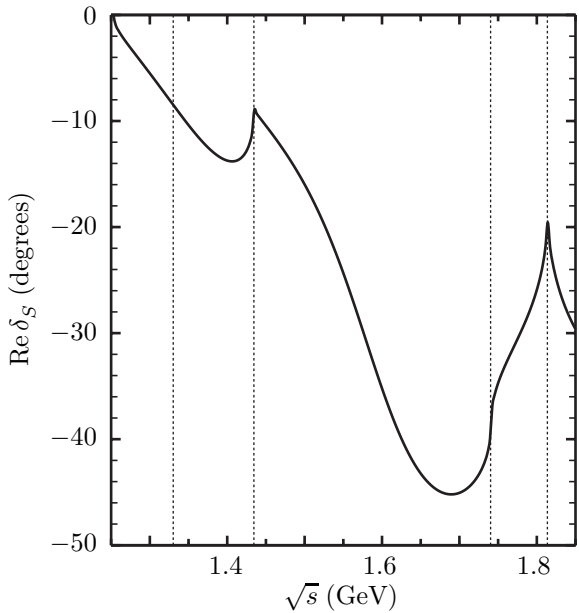

The lowest-order $S$-wave amplitude for  
$\,B\phi\rightarrow B'\phi'\,$  
with  $\,S=-1\,$  and  $\,I=1\,$  is derived from the first 
diagram in Fig.~\ref{diagrams,p1} and, in the CM frame, 
has the form 
\begin{eqnarray}   
f_{B\phi,B'\phi'}^{(S)}  \;=\;  
-C_{B\phi,B'\phi'}^{}\; \sqrt{m_B^{} m_{B'}^{}}\;   
{E_\phi^{}+E_{\phi'}^{}\over 16\pi f_\pi^2\sqrt{s}}   \;,  
\end{eqnarray}      
where  $\,C_{B'\phi',B\phi}^{}=C_{B\phi,B'\phi'}^{}.\,$  
Using the isospin states above, one obtains  
\begin{eqnarray}   
\begin{array}{c}   \displaystyle      
C_{\Lambda\pi,\Lambda\pi}^{}  \;=\;  
C_{\Lambda\pi,\Sigma\pi}^{}  \;=\;  0   \;,  
\hspace{3em}   
C_{\Lambda\pi,N\bar{K}}^{}  \;=\;  \sqrt{\mbox{$3\over 2$}}   \;,  
\hspace{3em}   
C_{\Lambda\pi,\Sigma\eta}^{}  \;=\;  0   \;,  
\hspace{3em}   
C_{\Lambda\pi,\Xi K}^{}  \;=\;  \sqrt{\mbox{$3\over 2$}}   \;,  
\vspace{1ex} \\   \displaystyle   
C_{\Sigma\pi,\Sigma\pi}^{}  \;=\;  -2   \;,  
\hspace{3em}   
C_{\Sigma\pi,N\bar{K}}^{}  \;=\;  -1   \;,  
\hspace{3em}   
C_{\Sigma\pi,\Sigma\eta}^{}  \;=\;  0   \;,  
\hspace{3em}   
C_{\Sigma\pi,\Xi K}^{}   \;=\;  +1   \;,  
\vspace{1ex} \\   \displaystyle   
C_{N\bar{K},N\bar{K}}^{}  \;=\;  -1   \;,  
\hspace{3em}    
C_{N\bar{K},\Sigma\eta}^{}  \;=\;  \sqrt{\mbox{$3\over 2$}}   \;,  
\hspace{3em}   
C_{N\bar{K},\Xi K}^{}  \;=\;  0   \;,  
\vspace{1ex} \\   \displaystyle   
C_{\Sigma\eta,\Sigma\eta}^{}  \;=\;  0   \;,  
\hspace{3em}   
C_{\Sigma\eta,\Xi K}^{}  \;=\;  \sqrt{\mbox{$3\over 2$}}   \;,  
\hspace{3em} 
C_{\Xi K,\Xi K}^{}  \;=\;  -1   \;.  
\end{array}      
\end{eqnarray}      
The resulting  $K$-matrix is written as  
\begin{eqnarray}   
K  \;=\;  
\left( \begin{array}{cc}   \displaystyle
K_{\rm oo} & K_{\rm oc}  
\vspace{1ex} \\   \displaystyle   
K_{\rm co} & K_{\rm cc}  
\end{array} \right)   \;,   
\end{eqnarray}      
where, with  
$\,f_{B\phi,B'\phi'}^{}\equiv f_{B\phi,B'\phi'}^{(S)},\,$  
\begin{eqnarray}   
\begin{array}{c}   \displaystyle   
K_{\rm oo}^{}  \;=\;  f_{\Lambda\pi,\Lambda\pi}^{}   \;,   
\hspace{2em}   
K_{\rm co}^{}  \;=\;  K_{\rm oc}^{\rm T}  \;=\;  
\left( \begin{array}{c}   \displaystyle  
f_{\Lambda\pi,\Sigma\pi}^{}  \vspace{1ex} \\   \displaystyle   
f_{\Lambda\pi,N\bar{K}}^{}  \vspace{1ex} \\   \displaystyle   
f_{\Lambda\pi,\Sigma\eta}^{}  \vspace{1ex} \\   \displaystyle   
f_{\Lambda\pi,\Xi K}^{}  
\end{array} \right)   \;,   
\vspace{2ex} \\   \displaystyle   
K_{\rm cc}^{}  \;=\;  
\left( \begin{array}{cccc}   \displaystyle  
f_{\Sigma\pi,\Sigma\pi}^{} & f_{\Sigma\pi,N\bar{K}}^{} & 
f_{\Sigma\pi,\Sigma\eta}^{} &  f_{\Sigma\pi,\Xi K}^{}   
\vspace{1ex} \\   \displaystyle   
& f_{N\bar{K},N\bar{K}}^{} & f_{N\bar{K},\Sigma\eta}^{} & 
f_{N\bar{K},\Xi K}^{}  
\vspace{1ex} \\   \displaystyle   
& & f_{\Sigma\eta,\Sigma\eta}^{} & f_{\Sigma\eta,\Xi K}^{}  
\vspace{1ex} \\   \displaystyle   
& & & f_{\Xi K,\Xi K}^{}  
\end{array} \right)   
\;=\;  K_{\rm cc}^{\rm T}   \;,       
\end{array}   
\end{eqnarray}      
the subscripts o and c referring, respectively, to open and 
closed channels at  $\,\sqrt{s}=m_\Xi^{}.\,$   
The unitarized $S$-wave amplitude for  
$\,\Lambda\pi\rightarrow\Lambda\pi\,$  is then given 
by~\cite{pilkuhn}        
\begin{eqnarray}   \label{Too}
T_{\rm oo}^{}  \;=\;  
{ K_{\rm r}^{}  \over  
 1-{\rm i} q_{\rm o}^{}\, K_{\rm r}^{} }   
\;=\;  
{ {\rm e}^{2{\rm i}\delta_{S}^{}} - 1  \over  
 2{\rm i}\, \bigl| \mbox{\boldmath$k$}_{\Lambda\pi}^{} \bigr| }   \;, 
\end{eqnarray}      
where  
\begin{eqnarray}   
\begin{array}{c}   \displaystyle      
K_{\rm r}^{}  \;=\;  
K_{\rm oo}^{}  
+  {\rm i} K_{\rm oc}^{} 
 \bigl( 1-{\rm i} q_{\rm c}^{} K_{\rm cc}^{} \bigr) ^{-1} 
 q_{\rm c}^{} K_{\rm co}^{}   \;, 
\vspace{2ex} \\   \displaystyle   
q_{\rm o}^{}  \;=\;  q_{\Lambda\pi}^{}   \;,  
\hspace{3em}  
q_{\rm c}^{}  \;=\;   
{\rm diag} \left( 
q_{\Sigma\pi}^{}, q_{N\bar{K}}^{}, q_{\Sigma\eta}^{}, q_{\Xi K}^{}  
\right)   \;,        
\end{array}      
\end{eqnarray}      
with    
$\,q_{B\phi}^{}= \bigl| \mbox{\boldmath$k$}_{B\phi}^{} \bigr| ,\,$  
the magnitude of the particle three-momentum in 
$\,B\phi\rightarrow B\phi\,$  scattering in the CM frame.  
We note that  $T_{\rm oo}^{}$  not only satisfies elastic 
unitarity exactly, 
but also reproduces the lowest-order  $\chi$PT  amplitude  
$f_{\Lambda\pi,\Lambda\pi}^{}$  (which happens to vanish in 
the $S$-wave case) as the chiral limit is approached.  
We further note that the diagonal elements of  $q_{\rm c}^{}$  
are purely imaginary at $\,\sqrt{s}=m_\Xi^{},\,$  
their corresponding channels being below threshold.  
It  follows that the  $S$-wave phase shift in  
$\,\Lambda\pi\rightarrow\Lambda\pi\,$  
scattering at  $\,\sqrt{s}=m_\Xi^{}\,$  is calculated to be  
\begin{eqnarray}   
\delta_{S}^{}  \;=\;  
\tan^{-1} \bigl( q_{\Lambda\pi}^{} K_{\rm r}^{} \bigr)  
\;\simeq\;  -7.3^\circ    \;.   
\end{eqnarray}      
If we drop the heavier $\Sigma \eta$ and $\Xi K$ channels, 
the phase shift is reduced in size to  
$\,\delta_{S}^{} \simeq -2.8^\circ.\,$   
If only the  $\Xi K$  channel is dropped, we find instead 
$\,\delta_{S}^{} \simeq -3.6^\circ.\,$  
These numbers are consistent with the fact that the $\Lambda\pi$  
state has nonzero $S$-wave couplings at leading order only to 
the $N\bar{K}$  and  $\Xi K$ states.    
Interestingly, the last two numbers, 
$\delta_{S}^{}\sim -3^\circ,\,$ are similar to 
the $\delta_{S}^{}$  in  Eq.~(\ref{deltaSP}), calculated using  
$\chi$PT  at order  $p^2$  with the parameter values from 
Ref.~\cite{kw},  in which the heavier channels were 
not explicitly considered.    
In Fig.~\ref{delta_S} we show the real part of $\delta_{S}^{}$ 
(which becomes complex above the $\Sigma\pi$ threshold) 
as a function of the center-of-mass energy, 
with all the four inelastic channels contributing.

We remark here that we have not included contributions from   
the  $B\phi\phi^\prime$ states in our  $K$-matrix as they are 
not expected to be dominant, only entering at 
the two-loop level in~$\chi$PT. 
Similarly, with the lowest-order vertices that we have, there 
are no $S$-wave couplings to states with a decuplet baryon and 
a pseudo-scalar meson.

The leading-order $P$-wave amplitude for  
$\,B\phi\rightarrow B'\phi'\,$  
with  $\,S=-1\,$  and  $\,I=1\,$  is derived from the last 
three diagrams in Fig.~\ref{diagrams,p1}.  
In the CM frame,  its $\,J=\frac{1}{2}\,$  component can be 
written in the form   
\begin{eqnarray}   
f_{B\phi,B'\phi'}^{(P)}  \;=\;  
- D_{B\phi,B'\phi'}^{}\;     
{ |\mbox{\boldmath$k$}_\phi^{}| 
 |\mbox{\boldmath$k$}_{\phi'}^{}| \sqrt{m_B^{}m_{B'}^{}} 
 \over  8\pi f^2\sqrt{s} }   \;,  
\end{eqnarray}
where  $\,D_{B'\phi',B\phi}^{}=D_{B\phi,B'\phi'}^{},\,$  
and  $\mbox{\boldmath$k$}_\phi^{}$  
$\bigl(\mbox{\boldmath$k$}_{\phi'}^{}\bigr)$  
is the three-momentum of the initial (final) meson.  
We have collected in the Appendix the expressions for  
$D_{B\phi,B'\phi'}^{}$  corresponding to the five coupled 
channels. 
The  $P$-wave  $\,J=\frac{1}{2}\,$  phase-shift in  
$\,\Lambda\pi\rightarrow\Lambda\pi\,$  
scattering at  $\,\sqrt{s}=m_\Xi^{}\,$  is then 
calculated using the same method as in the $S$-wave case.  
The result for  $\,D=0.8,\,$  $\,F=0.5,\,$  and  
$\,{\cal C}=1.7\,$  is  
\begin{eqnarray}   
\delta_{P}^{}  \;\simeq\;  0.2^\circ   \;,   
\end{eqnarray}      
where isospin-symmetric masses have been used for 
the intermediate baryons.   
If chiral-symmetric masses  $\,m_{0}^{}\simeq 1.15\,\rm GeV$  
and  $m_T^{}\simeq 1.38\,\rm GeV\,$  are used for 
the intermediate baryons, the result is instead 
\begin{eqnarray}   
\delta_{P}^{}  \;\simeq\;  0.5^\circ   \;.   
\end{eqnarray}      
We have found that dropping one or more of the inelastic 
channels would not change these numbers dramatically in size, 
yielding a phase shift within the range  
$\,-1.5^\circ<\delta_{P}^{}<-0.4^\circ.\,$

As in the $S$-wave case, we have not included contributions    
from  the  $B\phi\phi^\prime$ states in our $P$-wave $K$-matrix.  
Neither have we considered states with only one decuplet 
baryon and no meson in the $s$-channel as they have  
$\,J=\frac{3}{2},\,$  but they were included as intermediate 
states in the  $\,J=\frac{1}{2}\,$  $u$-channel.  
Beyond the simplest approximation that we have made, one could 
add contributions from heavier states, such as those with one 
decuplet baryon and one pseudo-scalar meson, as well as those 
with heavier resonances.  
The neglect of heavier states is taken as part of 
the uncertainty of our estimate.

\section{Conclusion}
   
We have studied the  $\Lambda\pi$ scattering phase shifts at 
$\,\sqrt{s}=m_\Xi^{}\,$  beyond leading order in chiral 
perturbation theory.    
With next-to-leading-order $\chi$PT, we find results that 
are consistent within factors of two with the lowest-order 
phase shifts.   
Within a  $K$-matrix approach, we find that   
unitarity effects from coupled channels enhance 
the $S$-wave phase shift, which could be as large as 
$\,-7^\circ,\,$  but they do not change the $P$-wave phase 
shift significantly.   
The large $S$-wave value is driven mostly by couplings to 
the heavier channels.  
Since the two approaches do not incorporate exactly the same 
physics, their results may be combined to indicate that   
$\,-7.3^\circ<\delta_{S}^{}<+0.4^\circ\,$  and   
$\,-3.5^\circ<\delta_{P}^{}<+0.5^\circ,\,$  
leading to  
$\,-7.8^\circ<\delta_{S}^{}-\delta_{P}^{}<+3.9^\circ.\,$   
Our results also indicate that more-refined future 
calculations with~$\chi$PT as a starting point 
should include the effects of coupled channels,   
as has been done in the~$N\bar{K}$  case~\cite{kw,or,mo}.       
Finally, it is possible to extract these phase shifts from 
experiment.  
We expect E871 to present results in the near future 
from their analysis of polarization in  
$\,\Xi \rightarrow \Lambda \pi\,$ decays.

\bigskip\bigskip  
   
\noindent 
{\bf Acknowledgments}$\,$    
The work of J.T. and  G.V. was supported in part by DOE under 
contract number DE-FG02-92ER40730.  
The work of J.T. was also supported by DOE under 
contract number DE-FG02-96ER40989. This work was also supported in part
by the Australian Research Council.
G.V. thanks the Special Research Centre for the Subatomic 
Structure of Matter at the University of Adelaide 
for their hospitality and partial support.  
We thank  S.~Gardner, K.B.~Luk, S.~Pakvasa, and  M.J.~Savage  
for discussions.

\bigskip\bigskip  

\noindent
{\Large\bf Appendix}

For the five coupled channels considered here, one obtains    
\begin{eqnarray}   
\begin{array}{c}   \displaystyle      
D_{\Lambda\pi,\Lambda\pi}^{}  \;=\;  
{\frac{2}{3} D^2\over \sqrt{s}-m_\Sigma^{}}
- {\frac{2}{9} D^2\over E_\Lambda^{}-E_\pi^{}-m_\Sigma^{}}   
+ { \frac{2}{9} {\cal C}^2  \over  
   E_\Lambda^{}-E_\pi^{}-m_{\Sigma^*}^{} }   \;,   
\vspace{2ex} \\   \displaystyle   
D_{\Lambda\pi,\Sigma\pi}^{}  \;=\;  
{\frac{4}{\sqrt{6}} D F\over\sqrt{s}-m_\Sigma^{}}
+ {\frac{4}{3\sqrt{6}} D F\over E_\Lambda^{}-E_\pi'-m_\Sigma^{}}   
+ { \frac{4}{9\sqrt{6}} {\cal C}^2  \over  
   E_\Lambda^{}-E_\pi'-m_{\Sigma^*}^{} }   \;,   
\vspace{2ex} \\   \displaystyle   
D_{\Lambda\pi,N\bar{K}}^{}  \;=\;  
{-\sqrt{\frac{2}{3}} D(D-F)\over\sqrt{s}-m_\Sigma^{}}  
- { \frac{1}{3\sqrt{6}} \bigl( D^2+4D F+3F^2 \bigr)  \over  
   E_\Lambda^{}-E_K'-m_N }   \;,  
\vspace{2ex} \\   \displaystyle   
D_{\Lambda\pi,\Sigma\eta}^{}  \;=\;  
{\frac{2}{3} D^2\over\sqrt{s}-m_\Sigma^{}}  
+ { \frac{2}{9} D^2\over E_\Lambda^{}-E_\eta'-m_\Lambda }   \;,  
\vspace{2ex} \\   \displaystyle   
D_{\Lambda\pi,\Xi K}^{}  \;=\;  
{-\sqrt{\frac{2}{3}} D(D+F)\over\sqrt{s}-m_\Sigma^{}}
- { \frac{1}{3\sqrt{6}} \bigl( D^2-4D F+3 F^2 \bigr)  \over 
   E_\Lambda^{}-E_K'-m_\Xi^{} }   
- { \frac{4}{9\sqrt{6}} {\cal C}^2  \over  
   E_\Lambda^{}-E_K'-m_{\Xi^*}^{} }   \;,   
\end{array}  
\end{eqnarray}   
\begin{eqnarray}   
\begin{array}{c}   \displaystyle  
D_{\Sigma\pi,\Sigma\pi}^{}  \;=\;  
{4 F^2\over \sqrt{s}-m_\Sigma^{}}
+ {\frac{2}{9} D^2\over E_\Sigma^{}-E_\pi'-m_\Lambda^{}}   
- {\frac{2}{3} F^2\over E_\Sigma^{}-E_\pi'-m_\Sigma^{} }   
+ { \frac{2}{27} {\cal C}^2  \over  
   E_\Sigma^{}-E_\pi'-m_{\Sigma^*}^{} }   \;,   
\vspace{2ex} \\   \displaystyle   
D_{\Sigma\pi,N\bar{K}}^{}  \;=\;  
{-2(D-F)F\over\sqrt{s}-m_\Sigma^{}}   
+ { \frac{1}{3} \bigl( D^2-F^2 \bigr)  \over  
   E_\Sigma^{}-E_K'-m_N^{} }   
- {\frac{8}{27} {\cal C}^2\over E_\Sigma^{}-E_K'-m_{\Delta}^{}}   \;,   
\vspace{2ex} \\   \displaystyle   
D_{\Sigma\pi,\Sigma\eta}^{}  \;=\;  
{\frac{4}{\sqrt{6}} D F\over\sqrt{s}-m_\Sigma^{}}   
- {\frac{4}{3\sqrt{6}} D F\over E_\Sigma^{}-E_\eta'-m_\Sigma^{}}   
+ { \frac{4}{9\sqrt{6}} {\cal C}^2  \over  
   E_\Sigma^{}-E_\eta'-m_{\Sigma^*}^{} }   \;,   
\vspace{2ex} \\   \displaystyle   
D_{\Sigma\pi,\Xi K}^{}   \;=\;  
{-2 (D+F)F\over\sqrt{s}-m_\Sigma^{}}   
- { \frac{1}{3} \bigl( D^2-F^2 \bigr)  \over  
   E_\Sigma^{}-E_K'-m_\Xi^{}}   
- {\frac{4}{27} {\cal C}^2\over E_\Sigma^{}-E_K'-m_{\Xi^*}^{}}   \;,   
\end{array}      
\end{eqnarray}      
\begin{eqnarray}   
\begin{array}{c}   \displaystyle  
D_{N\bar{K},N\bar{K}}^{}  \;=\;  
{(D-F)^2\over\sqrt{s}-m_\Sigma^{}}   \;,  
\hspace{3em}  
D_{N\bar{K},\Sigma\eta}^{}  \;=\;  
{-\frac{\sqrt{6}}{3} D(D-F)\over\sqrt{s}-m_\Sigma^{}}   
- { \frac{\sqrt{6}}{18} \bigl( D^2-4D F+3F^2 \bigr)  \over  
   E_N^{}-E_\eta'-m_N^{}}   \;,   
\vspace{2ex} \\   \displaystyle   
D_{N\bar{K},\Xi K}^{}  \;=\;  
{D^2-F^2\over\sqrt{s}-m_\Sigma^{}}   
- { \frac{1}{18} \bigl( D^2-9 F^2 \bigr)  \over 
   E_N-E_K'-m_\Lambda^{} }   
+ { \frac{1}{6} \bigl( D^2-F^2 \bigr) \over  
   E_N-E_K'-m_\Sigma^{} }   
+ {\frac{2}{27} {\cal C}^2\over E_N^{}-E_K'-m_{\Sigma^*}^{}}   \;,   
\end{array}      
\end{eqnarray}      
\begin{eqnarray}   
\begin{array}{c}   \displaystyle  
D_{\Sigma\eta,\Sigma\eta}^{}  \;=\;  
{\frac{2}{3} D^2\over \sqrt{s}-m_\Sigma^{}}   
- {\frac{2}{9} D^2\over E_\Sigma-E_\eta'-m_\Sigma^{}}   
+ { \frac{2}{9} {\cal C}^2  \over  
   E_\Sigma^{}-E_\eta'-m_{\Sigma^*}^{} }   \;,   
\vspace{2ex} \\   \displaystyle   
D_{\Sigma\eta,\Xi K}^{}  \;=\;  
{-\frac{\sqrt{6}}{3} D(D+F)\over\sqrt{s}-m_\Sigma^{}}   
- { \frac{\sqrt{6}}{18} \bigl( D^2+4 D F+3 F^2 \bigr)  \over 
   E_\Sigma-E_K'-m_\Xi^{} }   
- { \frac{2\sqrt{6}}{27} {\cal C}^2  \over  
   E_\Sigma^{}-E_K'-m_{\Xi^*}^{} }   \;,   
\end{array}      
\end{eqnarray}      
\begin{eqnarray}   
\begin{array}{c}   \displaystyle  
D_{\Xi K,\Xi K}^{}  \;=\;  
{D^2+2 D F+F^2\over\sqrt{s}-m_\Sigma^{}}   
+ {\frac{4}{9} {\cal C}^2\over E_\Xi^{}-E_K'-m_{\Omega}^{}}   \;,    
\end{array}      
\end{eqnarray}      
where  $\,\sqrt{s}=E_B^{}+E_\phi^{}\,$   and  $E_\phi'$  
is the energy of  $\phi$  in the final state.

\end{document}